\documentclass[acus]{JAC2000}

\usepackage{graphicx}

\begin{document}
\title{\flushright{PSN THBT003}\\[15pt] \centering THE CDF-II ONLINE
SILICON VERTEX TRACKER}

\author{A.~Bardi, A.~Belloni, R.~Carosi, A.~Cerri, G.~Chlachidze, 
M.~Dell'Orso, S.~Donati\thanks{corresponding author}, \\ 
S.~Galeotti, P.~Giannetti, V.~Glagolev,
E.~Meschi, F.~Morsani, D.~Passuello, G.~Punzi, \\
L.~Ristori, A.~Semenov, F.~Spinella,\\ 
INFN, University and Scuola Normale 
Superiore of Pisa, I-56100 Pisa, Italy \\
A.~Barchiesi, M.~Rescigno, S.~Sarkar, L.~Zanello,\\
INFN Sezione di Roma I and University La Sapienza, 
I-00173 Roma, Italy\\
M.~Bari, S.~Belforte, A.M.~Zanetti,\\
INFN Sezione di Trieste I-34012 Trieste, Italy\\
I.~Fiori, \\
INFN and University of Padova I-35031 Padova, Italy\\
B.~Ashmanskas, M.~Baumgart, J.~Berryhill, M.~Bogdan,
R.~Culbertson, H.~Frisch, \\
T.~Nakaya, H.~Sanders, M.~Shochet, U.~Yang,\\
Enrico Fermi Institute, University of Chicago, Chicago, IL 60637, USA\\
Y.~Liu, L.~Moneta, T.~Speer, X.~Wu,\\
University of Geneve, CH-1211 Geneve 4, Switzerland}

\maketitle

\begin{abstract}
The Online Silicon Vertex Tracker is the new CDF-II 
level 2 trigger processor designed to reconstruct 
2-D tracks within the Silicon Vertex Detector with 
high speed and accuracy. By performing a precise
measurement of impact parameters the SVT allows 
tagging online B events which typically show 
displaced secondary vertices. 
Physics simulations show that this will 
greatly enhance the CDF-II B-physics capability.
The SVT has been fully assembled and operational 
since the beginning of Tevatron RunII in April 2001.
In this paper we briefly review the SVT design and 
physics motivation and then describe its performance 
during the early phase (April-October 2001) of run II. 
\end{abstract}

\section{INTRODUCTION}
CDF (Collider Detector at Fermilab) is a general purpose detector 
designed to study the high energy $p{\bar p}$ interactions produced 
at the Tevatron Collider. 
The Tevatron has recently completed major upgrades to achieve 
higher energy (990~GeV per beam) and instantaneous luminosity
($10^{32}~cm^{-2}s^{-1}$). 
CDF has been upgraded to CDF-II to cope with the higher 
interaction rate as well as to extend its physics reach 
in this new and more difficult environment\cite{Spalding}.

The CDF upgrades relevant for this paper are the tracker and 
the trigger. The tracker is composed of the Central Drift 
Chamber (COT) and three independent silicon detectors 
(Layer00, Silicon Vertex Detector and Intermediate Silicon
Layers detector).
The Layer00 is made of one layer of radiation hard silicon 
microstrips (with $r-\phi$ readout) placed just outside of 
the beam pipe ($r=1.5~$cm). The new Silicon Vertex Detector 
(SVXII)~\cite{Saverio,Franz} is made of five double-sided
(with $r-\phi$ and $r-z$ readout) microstrip sensors 
arranged in a 12-fold azimuthal geometry and segmented 
in 6 longitudinal barrels (3 mechanical units, each read 
out at both ends) along the beam line (CDF $z$-axis). 
The silicon layers are posed between 2.5 and 10.6~cm from 
the beamline. With the length of about 1 meter the
SVXII covers 2 units in pseudorapidity. The ISL is 
located between the SVXII and the COT with one central 
layer at 23 cm from the beamline and two forward/backward
layers respectively at 20 and 29 cm from the beamline.
The COT is a multilayer drift chamber which covers
the region between 46 cm and 131 cm from the beamline.
It provides a high resolution measurement of the curvature
and azimutal angle of the charged tracks.

The CDF trigger has been completely rebuilt and it consists 
of three levels. Its challenging task is to reduce the 5~MHz 
rate input to level 1 to the 50~Hz allowed as maximum output 
rate from level~3 which is written directly to tape. 
Levels 1 and 2 are implemented in hardware while level 3 is 
an executable running on a PC farm.
The level 1 device most relevant to this paper is the eXtremely 
Fast Track finder processor (XFT)~\cite{XFT} which reconstructs 
2-D tracks (in the $r-\phi$ plane, transverse to the beam line) 
in the central drift chamber (COT). The Online Silicon Vertex 
Tracker (SVT) is part of the level 2 trigger. It receives the 
list of COT tracks reconstructed by the XFT processor (for each 
track the curvature $c$ and azimutal angle $\phi$ are measured) 
and the digitized pulse heights on the silicon layers ($\sim 10^5$ 
channels). The SVT links the XFT tracks to the silicon hits
and reconstructs tracks with offline-like quality.
From simulation of the tracking algorithm on run~I data 
the expected resolution of the SVT is $\delta \phi \simeq 1~$mrad, 
$\delta P_t~\simeq~0.003 \cdot {P_t}^2$~GeV/c, $\delta d \simeq 35~\mu m$ 
($d$ is the track impact parameter, i.e. the distance of 
closest approach of the particle trajectory elix to the 
$z$-axis of the CDF reference system).

By providing a precision measurement of the impact parameter
of charged particle tracks SVT allows triggering on events 
containing long lived particles. B hadrons in particular have 
a decay lenght of the order of 500~micron and tracks which 
come out of the B decay vertices have an impact parameter 
on average greater than 100~micron. The opportunity offered
by the SVT of triggering directly on B hadron decay vertices 
is available for the first time at a hadron collider. It
greatly improves the capability of selecting online B events
which was tightly bound to leptonic decay modes in the
past. Given the very large cross section for producing B
hadrons of the Tevatron (100 microbarn), CDF-II expects 
to have access even to rare purely hadronic B decays like 
$B^\circ \rightarrow \pi \pi$ and 
$B_s \rightarrow D_s \pi \rightarrow$ hadrons, 
which are extremely interesting respectively for CP 
violation and $B_s$ mixing measurements.

\section{SVT working principle}

The SVT has a very short time to keep up with 
the 50~KHz level 1 accept rate and to perform 
its task (on average about 20~$\mu$s 
per event). For this reason the whole SVT design 
has been concentrated on speeding up operations.
The SVT has a widely parallelized design: it is 
made of 12 identical azimuthal slices (``wedges'') 
working in parallel. Each wedge receives and
processes data from only one SVXII 30$^{\circ}$
wedge. In addition the SVT reconstructs only
tracks in the transverse plane to the beamline
(stereo info from SVXII is dropped) and only 
with $p_t$ above 2~GeV/c.

The tracking process is performed in two steps:
\begin{itemize}
\item{Pattern recognition: candidate tracks 
are searched among a list of precalculated low 
resolution patterns (``roads'');}
\item{Track fitting: a full resolution fit of
the hit coordinates found within each road 
is performed using a linearized algorithm};
\end{itemize}

The pattern recognition step is performed in a completely 
parallel way by the Associative Memory system which uses 
full custom VLSI chips (AMchips \cite{AM}).
The AM system compares all the silicon clusters and XFT
tracks with the set of precalculated patterns. A pattern 
is defined as a combination of five bins (``SuperStrips''):
four SuperStrips correspond to the position coordinates of 
the particle trajectory on four silicon layers, which can 
be chosen among the five SVXII layers and Layer00, the fifth 
SuperStrip corresponds to the azimuthal angle of the particle 
trajectory at a distance of 12~cm from the beam line.
The output of AM system is the list of patterns (``roads'')
for which at least one hit has been found on each SuperStrip.
Each SVT wedge uses 32K patterns which cover more than 95\% 
of the phase space for $p_t \geq$ 2~GeV/c.
Simulation studies have shown that SVT performance is optimized 
(in terms of processing time and final resolution) by choosing a 
SuperStrip size of about 250~micron on the silicon layers and 
5$^\circ$ for the $\phi$ angle measured by XFT.

The track fitting method is based on linear approximations 
and principal component analysis~\cite{RobertoIEEE}.
The analytical relationship between the track parameters 
and the six measured hit coordinates (hit positions on four 
silicon layers, curvature, and azimuthal angle of XFT track) 
can be expressed in terms of six equations:
\begin{equation}
P_j = P_j({\bf x}) = {\bf F}_j \cdot {\bf x} + Q_j
\label{eq:linexp}
\end{equation}
where ${\bf x}$ is the vector of hit coordinates and 
${\bf P}$ = ($d$,c, $\phi$, $\chi_1$, $\chi_2$, $\chi_3$) 
are respectively the impact parameter ($d$), the curvature 
($c$) and the azimuthal angle ($\phi$) at the point 
of minimum approach to the $z$-axis, while 
$\chi_1$, $\chi_2$ and $\chi_3$ are three independent 
constraints which all real tracks must satisfy 
(within detector resolution effects).
${\bf F}_j$ and $Q_j$ are constants which depend only 
on the detector geometry and the magnetic field.

After pattern recognition has been performed each track 
candidate is confined within one road and this information 
can be used to simplify the computation of 
equation~\ref{eq:linexp}. The hit coordinates and 
track parameters are given by the sum of a term
which depends only on the SuperStrip edge ($\bf x_0$)
and an additional term ($\bf \delta x$) which depends on the 
posistion of the hits within the SuperStrips.
Equation~\ref{eq:linexp} becomes:
\begin{equation}
P_{0j} + \delta P_j = {\bf F}_j \cdot ({\bf x_0} + \delta {\bf x}) + Q_j
\label{eq:toedge}
\end{equation}
where, $P_{0j} = {\bf F}_j \cdot {\bf x_0} + Q_j$ is a constant 
which depends on the single road, while $ \delta P_j$ is the 
correction which depends on the precise hit positions inside 
the road. The $P_{0j}$ coefficients are calculated offline, 
stored in RAMs and used on a track-by-track basis. Therefore 
the track fitting task reduces to the fast computation of 
simple scalar products (done by FPGA chips) 
\begin{equation}
\delta P_j = {\bf F}_j \cdot {\bf \delta x}.
\label{eq:fpga}
\end{equation}
which determine the small correction to add to the road 
dependent constant terms. The SVT output is the list of 
high precision tracks which is sent to the trigger 
processors for the final level 2 trigger decision.

SVT is made by over one hundred VME boards housed in 
eight crates~\cite{SimoneFI}. The installation has been 
completed and the system has been fully operational since 
the beginning of 2001.
In the following section we report on the performance 
achieved by the system in the early phase (April-October
2001) of run II.

\section{SVT performance}

\begin{figure}[t]
\begin{center}
\includegraphics*[width=75mm]{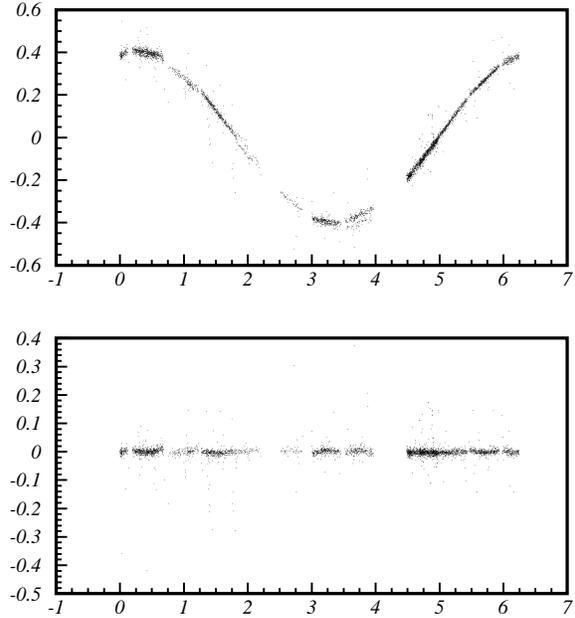}
\caption{
$d - \phi$ correlation for SVT track candidates 
before ($top$) and after correcting for beam offset 
($bottom$). The average beam position in the transverse 
plane is measured with a fit to the top plot ($x_0$=0.0995 cm 
and $y_0$=-0.3895~cm) with a precision of $\simeq 3$~micron. 
The regions without points around $\phi$=2.2 and 
$\phi$=4.2~rad are due to SVXII wedges which were turned off
when the data analysed in this paper were taken.  
The scale on the $x$ axis is radians, the scale on the $y$ 
axis is cm.
\label{figura1}}
\end{center}
\end{figure}

The correlation of the impact parameter $d$ versus 
the azimuthal angle $\phi$ of candidate tracks 
(Figure~\ref{figura1} ({\em top})) is the first 
evidence that the SVT tracks are good. The plot
shows the typical sine wave shape which is due
to the fact that the position of the interaction 
vertex in the transverse plane ($x_0$, $y_0$) is
displaced from the origin of CDF reference system.
In this case the relationship between $d$ and $\phi$ 
for primary tracks is:
\begin{equation}
d = - x_0 sin(\phi) + y_0 cos(\phi).
\label{eq:dfcorr}
\end{equation}
A fit to the $d-\phi $ scatter plot provides a measure 
of the ($x_0$, $y_0$) coordinates with an accuracy of 
few microns.

\begin{figure}[t]
\begin{center}
\includegraphics*[width=75mm]{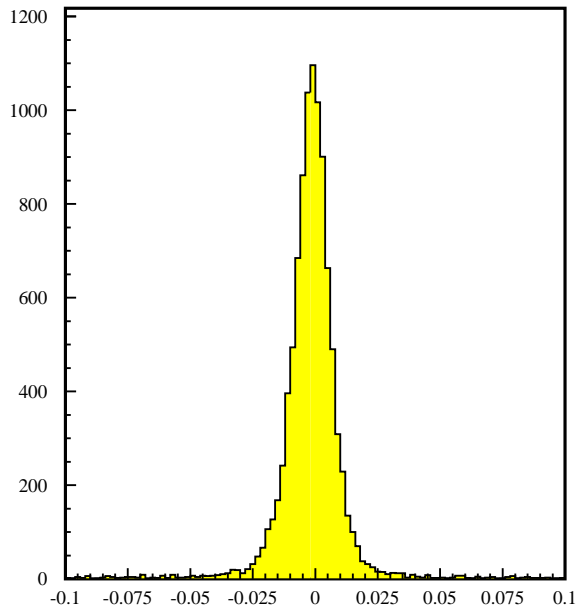}
\caption{Impact parameter distribution (scale 
is cm) after correcting for the beam offset.
\label{figura2}}
\end{center}
\end{figure}

A beam displacement of few millimiters is a typical 
running condition for CDF. In principle this condition
could be a problem because it generates unphysical 
large impact parameters which are erroneously interpreted 
by the level 2 trigger as the presence of B decays. 
In practice this potential problem is avoided by having
a process running on the SVT VME crate controller and 
performing a fit of the $d-\phi$ correlation to determine
the beam offset. The beam position parameters are 
transferred to the SVT which subtractes the beam offset
internally (Figure~\ref{figura1} ({\em bottom})). To
remove possible additional misalignments among SVXII
barrels, this is done independently for tracks in each 
of the six SVXII $z$-barrels. Following this procedure
the level 2 trigger receives physical impact parameters
(i.e. measured with respect to the actual beam position).
Figure~\ref{figura2} shows the impact parameter distribution 
after correcting for the beam offset. This distribution 
is given by the convolution of the actual transverse beam 
profile with the intrinsic impact parameter resolution.
A gaussian fit gives $\sigma \simeq 69$~micron.

A big effort has been put in understanding the various 
factors which contribute to the SVT impact parameter 
resolution. The goal is to improve it as much as possible,
because the worse is the resolution the higher is the 
rate of the triggers based on impact parameter cuts.
The potentially most relevant contribution to the $d$ 
resolution arises from the $z$-misalignment between the 
detector and the beam. The effect of a $z$-misalignment 
($mx=\delta x/ \delta z$, $my=\delta y/ \delta z$) 
shows up as a residual modulation in the impact parameter 
$d - \phi$ correlation obtained after correcting for the 
beam offset:
\begin{equation}
d^\prime = - m_x z_0 sin(\phi) + m_y z_0 cos(\phi)
\label{eq:dfz}
\end{equation}
where $z_0$ is the $z$-coordinate of the point of closest 
approach of the particle trajectory to the $z$-axis.
Unfortunately $z_0$ is not measured by the SVT, which 
receives the only $z$ information from the SVXII segmentation 
into six barrels, therefore the beam tilt results in 
an irreducible widening of the $d$ distribution.
In order to make this spread small compared to the natural 
beam width, SVT requires the detector and the beam to be  
parallel within 100~$\mu$rad. Assembly of the SVXII barrels 
indeed met this specification~\cite{Franz}, The alignement 
of the beam orbit turned out to be more challenging.
During the April-October data taking period, the beam slope 
was significantly large: $m_x \simeq 600~\mu$rad, 
$m_y \simeq 150~\mu$rad, well beyond the SVT specification.
The effect of the $z$-misalignment on the impact parameter 
distribution was estimated using a special run taken with 
an approximately null beam tilt. In this case the $d$ 
distribution had $\sigma \simeq 59$~micron, which is 10 
micron smaller than the width found for the runs taken
with significant beam tilt. Work is in progress in order
to have a good alignment as the standard running 
condition.

There are two additional major contributions affecting 
the impact parameter resolution. The first is the 
relative misalignment among SVXII wedges. This 
contribution can be easily corrected by performing 
the beam position fit and the corresponding impact 
parameter beam offset subtraction independently in
each wedge. The size of this correction to the 
resolution is approximately 6~micron.
The second contribution to the impact parameter
resolution is a consequence of the linear approximation 
of the SVT track fitting method, which assumes a first 
order expansion centered at the beam position.
Since during the April-October data taking the beam was 
very far from its nominal position ($\simeq 4~$mm away) 
the effect of non linearity was significant: it reduced 
the impact parameter resolution by approximately 5 micron.
This effect can be corrected by recalculating the
constants used in equation~\ref{eq:toedge} centering the
expansion on the measured beam position. After applying
all these corrections the impact parameter distribution
was found to have a gaussian shape with a sigma of 48 
micron for a run taken with the beam aligned in $z$
\footnote{There are additional effects:  
a residual non linearity in $d$ and $\phi$, and the 
misalignement of silicon layers within a wedge. The 
correction for these effects is less straightforward 
because it requires reprogramming some SVT boards
and adjusting the detector geometry of the SVT maps.
The effect of these additional correction has been 
studied with the simulation and reduces the the width 
of $d$ distribution to 45 micron.}
.All these corrections can be easily implemented in the 
SVT by modifying the constans loaded on the boards. 
They will be implemented in the next data taking 
after the October shutdown. The only ingredient out of 
the SVT control is the beam alignment along $z$.

As anticipated the impact parameter distribution 
is the convolution of the actual transverse beam
profile with the intrinsic impact parameter resolution.
By using equation \ref{eq:dfcorr} it can be shown that
the covariance of the impact parameters of two tracks
originating from the same vertex is proportional
to the cosine of the opening angle between the two
tracks:
\begin{equation}
\sigma_{d_1 d_2} = < d_1 \cdot d_2 > = \sigma^2_B\cdot\cos(\Delta\phi)
\label{eq:d1d2}
\end{equation}
where the constant $\sigma_B$ is the width of the 
actual transverse beam profile. By using this relation
and October 2001 data it has been possible to measure
the beam width (33 micron) and consequently determine 
the intrinsic impact parameter resolution of the SVT 
(35 micron) by subtracting in quadrature $\sigma_B$
from the impact parameter distribution width (48 micron).
The estimate for the SVT impact parameter resolution
is well in agreement with early SVT performance
simulations.

\section{Physics prospects with the SVT}
CDF has designed a trigger strategy (``hadronic B trigger'') 
based only on tracking to select hadronic B decays interesting
for CP violation and B$_s$ mixing. The idea is to
select charmless decays (B$^0_d \rightarrow h^+ h^-$)
for CP violation and a collection of charmed decays
(B$^0_s \rightarrow D^-_s \pi^+$ and 
B$^0_s \rightarrow D^-_s \pi^+\pi^-\pi^+$ with the
D$^-_s \rightarrow \phi \pi^-$ and 
D$^-_s \rightarrow K^{*\circ} K^-$) for mixing.
This trigger strategy relies on XFT tracks at level 1 
and SVT tracks at level 2. The most important ingredient
are the level 2 impact parameter cuts which reduce the 
trigger rate by three orders of magnitude. For the
success of this trigger strategy the SVT performance 
is essential.

In October 2001 CDF took the first test runs which implemented 
a simplified version of the hadronic B trigger. The level 1 
required two opposite charge 
XFT tracks with $p_t >$ 2 GeV/c, $p_{t1}+p_{t2} >$ 5.5 GeV/c
and an opening angle smaller than 135$^{\circ}$. The level 2
required two good SVT tracks with $p_t >$ 2 GeV/c (no impact
parameter threshold was set for these preliminary test runs).
Using this data ($\simeq$ 15~nb$^{-1}$) and a very simple
selection based on decay length cuts, a small signal of 
$D^\circ \rightarrow K \pi$ was reconstructed~\cite{RobertoIEEE}
(Figure~\ref{figura2}). Although this is not a fully reconstructed
hadronic B signal yet, it shows that the purity of the triggered
sample allows extracting some heavy flavour signal even from a
very small sample (15 nb$^{-1}$). As data taking proceeds we 
expect to perform more sophisticated analysis and to isolate
the first hadronic B signals. With the 2 fb$^{-1}$ of data
CDF-II plans to collect in run II we expect to contribute
significantly to the field of CP violation in B decays and
B$_s$ mixing.

\begin{figure}[b]
\begin{center}
\includegraphics*[width=75mm]{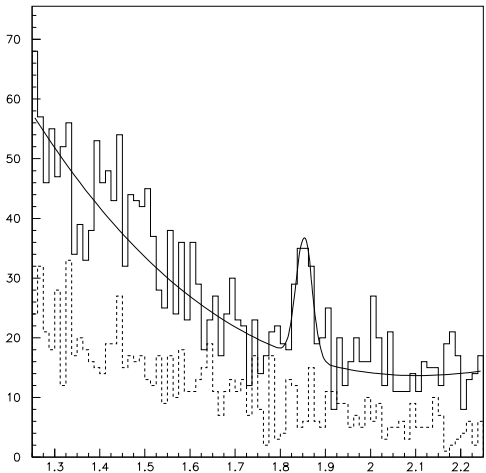}
\caption{
Invariant mass distribution for track pairs
assuming the two tracks to be $\pi$ and $K$, 
and applying cuts optimized to select a 
$D^\circ \rightarrow K \pi$ signal. The scale
of the horizontal axis is GeV/c$^2$.}
\label{figura4}
\end{center}
\end{figure}

\section{Conclusions}

The Online Silicon Vertex Tracker is the new level 2 trigger 
processor dedicated to the reconstruction of tracks within 
the tracking chamber and the silicon vertex detector of the 
CDF-II experiment. The SVT has been successfully installed 
and operated during the preliminary phase of run II. 
The performance of the device has already shown to be very 
close to the design. The impact parameter resolution is
as good as needed for a successful operation of the trigger
provided that the tilt between the beam and the detector is 
not larger than 100 microradiants.
CDF relies on SVT tracks for trigger strategies dedicated
to the selection of hadronic B decays. Preliminary test
runs implementing these strategies have been taken in 
October 2001 and the first signals of heavy flavours 
have been found. 
Improvements of the SVT performance are expected in the
near future as the operating conditions will evolve 
from the commissioning to a more stable phase.
Pedestal adjustments, dead channels suppressions and 
a better tuning of the clustering algorithm are expected.
Fine tuning of the SVT (corrections for non-linearities
and relative wedge-to-wedge misalignments) will soon
be implemented.

\end{document}